\documentclass[12pt]{article}  %fuer beidseitige Version
\usepackage[a4paper,textwidth=490.0pt,textheight=703.1pt]{geometry}
\usepackage{hyperref}
\usepackage{slashed}
\usepackage{graphicx}
\usepackage{epsfig}
\usepackage{amsmath}
\usepackage{amssymb}
\usepackage{ulem}
\usepackage{mdwlist}
\usepackage{color}                                                  
\usepackage{float}
\usepackage[rflt]{floatflt}
\usepackage{slashed}          
\usepackage{cite}
\usepackage{mathtools}
\usepackage{dsfont}
\usepackage{subfig}

\newcommand{\Atil}{\tilde{A}}
 \newcommand{\Ctil}{\tilde{C}}

\usepackage{array}

\usepackage[english]{babel}
\usepackage[latin1]{inputenc}
\usepackage[T1]{fontenc}
\usepackage{ae}

\usepackage{url}

\usepackage{amsmath, amsthm, amssymb}

\newtheorem{thm}{Theorem}[section]

\usepackage{array}

\usepackage[english]{babel}
\usepackage[latin1]{inputenc}
\usepackage[T1]{fontenc}
\usepackage{ae}

\newtheorem{definition}[thm]{Definition}

\usepackage{rotating}

\usepackage{graphicx}
\newcounter{mmacnt}
\def\restartmma{\setcounter{mmacnt}{0}}
\restartmma \catcode`|=\active
\def|#1|{\mathrm{#1}}
\catcode`|=12
\newenvironment{mma}{
 \par\smallskip
 \catcode`|=\active
 \parskip=0pt\parindent=0pt % locally
 \small
 \def\In##1\\{%
   \def\linebreak{\hfill\break\null\qquad}%
   \refstepcounter{mmacnt}
   \hangindent=2.5em\hangafter=0
   \leavevmode
   \llap{\tiny\sffamily In[\arabic{mmacnt}]:=\kern.5em}%
   \mathversion{bold}\footnotesize$\displaystyle##1$\normalsize
   \mathversion{normal}\par
 }%
 \def\Print##1\\{%
   \def\linebreak{\hfill\break}%
   \hangindent=2.5em\hangafter=0
   \leavevmode ##1\par}%
 \def\Out##1\\{%
   \def\linebreak{$\hfill\break\null\hfill$}%
   \kern\abovedisplayskip\par
   \hangindent=2.5em\hangafter=0
   \leavevmode
   \llap{\tiny\sffamily Out[\arabic{mmacnt}]=\kern.5em}
   \footnotesize$\displaystyle##1$\normalsize\hfill\null\par
   \kern\belowdisplayskip
 }%
 \def\Warning##1##2\\{%
   \def\linebreak{\hfill\break}%
   \hangindent=2.5em\hangafter=0
   \leavevmode
   {\scriptsize##1 : ##2}\par}%
}{%
 \par\smallskip
}

\allowdisplaybreaks[4]

\usepackage{color}

\newenvironment{fshaded}{%
\MakeFramed {\FrameRestore}
}%
{\endMakeFramed}

\usepackage{tikz}
\usetikzlibrary{matrix}

\allowdisplaybreaks[4]

\begin{document}
\setlength{\baselineskip}{0.515cm}
\sloppy
\thispagestyle{empty}
\tikzset{
	graviton/.style={decorate,line width=0.25mm, decoration={snake,amplitude=.5mm, segment length=2mm}},
	massive/.style={postaction={decorate},
		line width=0.4mm,
	},
}
 
\makeatletter
\def\simgt{\mathrel{\lower2.5pt\vbox{\lineskip=0pt\baselineskip=0pt
           \hbox{$>$}\hbox{$\sim$}}}}
\def\simlt{\mathrel{\lower2.5pt\vbox{\lineskip=0pt\baselineskip=0pt
           \hbox{$<$}\hbox{$\sim$}}}}
\makeatother
           
\def\draftnote#1{{\textcolor{red}{\it #1}}}

\def\FT#1{{\color{magenta} [FT: #1]}}
\def\DK#1{{\color{blue} [DK: #1]}}
\def\TS#1{{\color{cyan} [TS: #1]}}

\def\fig#1{Fig.~\ref{#1}}

\def\eqn#1{Eq.~\eqref{#1}}
\def\eqns#1.#2{Eqs.~\eqref{#1} and~\eqref{#2}}
\def\spa#1.#2{\left\langle#1\,#2\right\rangle}
\def\spb#1.#2{\left[#1\,#2\right]}
\def\sand#1.#2.#3{%
\left\langle#1{\vphantom1}\right|{#2}\left|#3\right]}%
\def\sandmp#1.#2.#3{%
\left\langle#1{\vphantom1}\right|{#2}\left|#3\right]}%
\def\sandpm#1.#2.#3{%
\left[#1{\vphantom1}\right|{#2}\left|#3\right\rangle}%
\def\sandmm#1.#2.#3{%
\left\langle#1{\vphantom1}\right|{#2}\left|#3\right\rangle}%
\def\sandpp#1.#2.#3{%
\left[#1{\vphantom1}\right|{#2}\left|#3\right]}%

\def\pp{\sigma}
\def\LL{\mathcal{L}}
\def\phis{\phi_{s}}
\def\CC{C_{2}}
\def\PLS{\mathbb{S}}
\def\Es{\mathcal{E}}
\def\gS{\mathsf{S}}
\def\oS{\mathsf{S}}
\def\sS{\mathsf{S}}
\def\order{O}
\def\op{\mathcal{O}}
\def\opK{\mathcal{K}}

\def\hdelta{{\hat\delta}}
\def\opa{{\hat a}}
\def\opT{{\mathbb{T}}}

%%% Hamiltonian
\def\opH{\mathcal{H}}
\def\opp{\bm p}
\def\opr{\bm r}
\def\opL{\left(\opr \times \opp\right)}
\newcommand{\spinStr}[1]{\Sigma_{#1}}
\def\clS{\textbf{S}}
\def\clK{\textbf{K}}
\def\KxS{X}
\def\clKxS{\textbf{\KxS}}
\def\eftSigma{\sigma}

\def\doe{\partial}
\def\bs{\boldsymbol}
\def\mc{\mathcal}
\def\clp{\bm p}
\def\clpb{\bar{\bm p}}
\def\clq{\bm q}

\newcommand{\sym}[1]{\{#1\}}

\def\nn{\nonumber}
\def\vs{\vskip 0cm }

\newcommand{\MeV}{\rm MeV}
\newcommand{\GeV}{\rm GeV}
\newcommand{\Section}[1]{\section{#1}}
\newcommand{\be}{\begin{equation}}
\newcommand{\ee}{\end{equation}}
\newcommand{\eq}[2]{\be\begin{aligned}#1 \label{#2}\end{aligned}\ee}

\newcommand{\Fig}[1]{Fig.~\ref{#1}}
\newcommand{\Eq}[1]{Eq.~\eqref{#1}}
\newcommand{\Eqs}[2]{Eqs.~\eqref{#1} and \eqref{#2}}
\newcommand{\Sec}[1]{Sec.~\ref{#1}}
\newcommand{\Secs}[2]{Secs.~\ref{#1} and \ref{#2}}
\newcommand{\App}[1]{App.~\ref{#1}}
\newcommand{\vev}[1]{\langle #1 \rangle}
\newcommand{\bra}[1]{\langle #1 |}
\newcommand{\ket}[1]{| #1 \rangle}

\newcommand{\sslash}[1]{\ensuremath\raisebox{-0.00cm}{{\small\slash}}\hspace{-0.21cm}#1\/}
\newcommand{\dd}[1]{\frac{\partial}{\partial #1}}

\newcommand{\OurOrder}{  {\cal{O}}(G^3)  }

\newcommand{\ECM}{E_{\rm CM}}
\newcommand{\pCM}{\bm{p}_{\rm CM}}

\newcommand{\mbf}[1]{\mathbf{#1}}

\newcommand{\AEFT}{A_{\rm EFT}}

\newcommand{\MEFT}{M_{\rm EFT}}

\newcommand{\II}{{\cal I}}

\newcommand{\E}{{\rm E}}
\newcommand{\K}{{\rm K}}

\newcommand{\gfkt}[3]{\ell_{#1,#2}^{(#3)}}
\newcommand{\pot}{\rm pot}

\renewcommand{\imath}{\mathrm{i}}

\def\topbotatom#1{\hbox{\hbox to 0pt{$#1\bot$\hss}$#1\top$}} \newcommand*{\topbot}{\mathrel{\mathchoice{\topbotatom\displaystyle} {\topbotatom\textstyle} {\topbotatom\scriptstyle} {\topbotatom\scriptscriptstyle}}}

\newcommand{\tabeq}[2]{ \parbox{#1}{  \be\begin{aligned}#2 \end{aligned} \nonumber \ee }}

%%%\title{{\sf \footnotesize 
\begin{flushleft}
DESY 25-127 \hfill     {\tt arXiv:2509.16124 [hep-ph]}\\ 
ZU-TH 59/25 \hfill September 2025 \\
MPP-2025-182 \\
RISC Report Series 25--08
\end{flushleft}

\vspace*{2cm}
\begin{center}
{\Large \bf The three-loop single-mass heavy-flavor corrections to the}

\vspace*{2mm}
{\Large \bf structure 
functions \boldmath $F_2(x,Q^2)$ and $g_1(x,Q^2)$}

\vspace*{30mm}
\normalsize
{J. Ablinger$^a$}, {A. Behring$^b$}, {J. Bl\"umlein$^{c,d}$}, {A. De Freitas$^a$},
{A. von Manteuffel$^e$}, 

\vspace*{2mm} 
{C. Schneider$^a$} and {K. Sch\"onwald$^{f,g}$}

\vspace*{5mm}
{\it $^a$~Johannes Kepler University Linz, Research Institute for Symbolic 
Computation (RISC), Altenberger Stra\ss{}e 69, A-4040, Linz, Austria}

\vspace*{2mm}
{\it $^b$~Max-Planck-Institut f\"ur Physik 
Boltzmannstra\ss{}e 8, 85748 Garching, Germany}

\vspace*{2mm}
{\it $^c$Deutsches Elektronen-Synchrotron DESY, Platanenallee 6, 15738 Zeuthen, Germany}

\vspace*{2mm}
{\it $^d$Institut f\"ur Theoretische Physik III, IV, TU Dortmund, Otto-Hahn 
Stra\ss{}e 4, 44227 Dortmund, Germany}

\vspace*{2mm}
{\it $^e$Institut f\"ur Theoretische Physik, Universit\"at Regensburg, 93040 
Regensburg, Germany}

\vspace*{2mm}
{\it $^f$
Physik-Institut, Universit\"at Z\"urich, Winterthurerstra\ss{}e 190, CH-8057 Z\"urich,
Switzerland}

\vspace*{2mm}
{\it $^g$ CERN, Theoretical Physics Department,
CH-1211 Geneva 23, Switzerland} 
\end{center}

\vfill
\begin{abstract}
  \noindent
  We present quantitative results on the single-mass heavy-flavor contributions in the
  region of large virtualities $Q^2$ up to three-loop order to the unpolarized structure 
  function $F_2(x,Q^2)$ and the polarized structure function $g_1(x,Q^2)$ for the first 
  time. These results are relevant for precision QCD analyses of the World deep-inelastic 
  data and the data taken at future colliders, such as the Electron--Ion Collider, since 
  the scaling violations due to massless and massive Wilson coefficients are significantly 
  different. In order to measure the strong coupling constant $\alpha_s(M_Z^2)$ and the 
  twist-2 parton distribution functions consistently at highest precision, the 
  next-to-next-to-leading order corrections have to be taken into account. Furthermore, 
  the complete three-loop corrections will allow to reduce the present theory error of the 
  charm mass $m_c$ as measured form deep-inelastic data. We provide a fast and precise 
  public numerical code for the unpolarized and polarized massive Wilson coefficients in 
  the asymptotic region.
\end{abstract}

%\pacs{12.38.Bx,12.38.Qk,14.65.Dw,14.65.Fy}
%\maketitle

%- }}}
%- {{{ Intro.:

\newpage
\section{Introduction}  
\label{sec:1}

\vspace*{1mm}
\noindent
The heavy-flavor corrections to deep-inelastic structure functions form an 
essential contribution
in the region of smaller values of Bjorken $x$. Their scaling violations are quite different from
those of the massless contributions, which requires their detailed knowledge in QCD precision 
analyses of these structure functions, such as the precision 
measurement of the strong coupling constant $a_s = \alpha_s/(4 \pi) = g_s^2/(16\pi^2)$ 
\cite{Bethke:2011tr,Moch:2014tta,Alekhin:2016evh,dEnterria:2022hzv}, the 
charm quark mass $m_c$ \cite{Alekhin:2012vu}, and the parton distribution functions, 
cf.,~e.g.,~Refs.~\cite{Accardi:2016ndt,Alekhin:2017kpj,PDF4LHCWorkingGroup:2022cjn}.

With the final results on the next-to-next-to-leading order (NNLO) single-mass corrections 
to the heavy-flavor Wilson coefficients in the region $Q^2 \gg m_Q^2$, where $m_Q$ is the 
heavy-quark mass, numerical predictions on the structure functions $F_2$ and $g_1$ can be 
made 
at this order for the first time.

As has been shown in Ref.~\cite{Alekhin:2012vu}, the extraction of the charm quark mass 
from DIS data relying on approximate representations \cite{Kawamura:2012cr} 
leads to a large theory uncertainty of
$\delta m_c^{\rm theory} = \tiny{\begin{array}{c} +0.00
\\ -0.07 \end{array}}~\GeV$  while the experimental error is $0.03~\GeV$. The 
experimental error will even decrease with new data from the
Electron-Ion Collider (EIC) in the future~\cite{Boer:2011fh,AbdulKhalek:2022hcn}.
Moreover, a series of Wilson coefficients, which are numerically smaller were not 
considered in this analysis. The completed NNLO calculation will lower this 
uncertainty. 

The purpose of this paper is to compile the single-mass three-loop contributions to    
the heavy-flavor Wilson coefficients to quantify the effect of the different 
contributions prior to analyses of experimental data. In the polarized case our results 
prepare for future precision measurements at the EIC.
These quantifications were not given before. Furthermore, a public numerical library
for the massive Wilson coefficients is provided. 

The massive Wilson coefficients can be represented by Mellin convolutions of the massive 
operator matrix elements 
(OMEs)  and pieces of the massless Wilson coefficients. 
The asymptotic single-mass Wilson coefficients are given by, cf.~Ref.~\cite{Bierenbaum:2009mv},
%---------------------------------------------------------------------------------
\begin{eqnarray}
\label{eqWIL1}
L_{2,q}^{\rm NS} &=&
a_s^2 \left[A_{qq,Q}^{{\rm NS}, (2)} +
\hat{C}^{{\sf NS}, (2)}_{2,q}\right]
+
a_s^3 \left[A_{qq,Q}^{{\rm NS}, (3)}
+  A_{qq,Q}^{{\rm NS}, (2)} C_{2,q}^{{\rm NS}, (1)}
+ \hat{C}^{{\rm NS}, (3)}_{2,q}\right],
\\
\label{eqWIL2}
{\tilde{L}}_{2,q}^{\rm PS} &=&
a_s^3 \left[~\Atil_{qq,Q}^{{\rm PS}, (3)}
+  A_{gq,Q}^{(2)}~~ \Ctil_{2,g}^{(1)}
+ \hat{\Ctil}^{{\rm PS}, (3)}_{2,q}\right],
\\
\label{eqWIL3}
{\tilde{L}}_{2,g} &=&
a_s^2 A_{gg,Q}^{(1)} \Ctil_{2,g}^{(1)}
+
a_s^3 \Bigl[~\Atil_{qg,Q}^{(3)}
+  A_{gg,Q}^{(1)}~~\Ctil_{2,g}^{(2)}
+  A_{gg,Q}^{(2)}~~\Ctil_{2,g}^{(1)}
+  ~A_{Qg}^{(1)}~~\Ctil_{2,q}^{{\rm PS}, (2)}
+ \hat{\Ctil}^{(3)}_{2,g}\Bigr],
\\
\label{eqWIL4}
H_{2,q}^{\rm PS}
&=& a_s^2 \left[~A_{Qq}^{{\sf PS}, (2)}
+~\Ctil_{2,q}^{{\sf PS}, (2)}\right]
+ a_s^3 \Bigl[~A_{Qq}^{{\rm PS}, (3)}
+~\Ctil_{2,q}^{{\rm PS}, (3)}
+ A_{gq,Q}^{(2)}~\Ctil_{2,g}^{(1)}
+ A_{Qq}^{{\rm PS},(2)}~C_{2,q}^{{\sf NS}, (1)}
\Bigr],
\\
H_{2,g} &=&  a_s   \left[A_{Qg}^{(1)} + \tilde{C}_{2,g}\right]
           + a_s^2    \left[A_{Qg}^{(2)} + A_{Qg}^{(1)} C_{2,q}^{{\rm NS},(1)} 
%\right.
%\nonumber\\ && \left.
+ A_{gg,Q}^{(1)} \tilde{C}_{2,g}^{(1)} + \tilde{C}_{2,g}^{(2)}\right]
+ a_s^3 \left[A_{Qg}^{(3)} + A_{Qg}^{(2)} C_{2,q}^{{\rm NS},(1)} \right.
\nonumber\\ && 
+ A_{gg,Q}^{(2)} \tilde{C}_{2,g}^{(1)}
+ A_{Qg}^{(1)}\left(C_{2,q}^{{\rm NS},(2)} + \tilde{C}_{2,q}^{{\rm PS},(2)}\right)
%\nonumber\\ && 
\left.
+ A_{gg,Q}^{(1)} \tilde{C}_{2,g}^{(2)} +\tilde{C}_{2,g}^{(3)}\right] 
\end{eqnarray}
%---------------------------------------------------------------------------------------------------------------
in Mellin space, where the Mellin convolutions simplify to ordinary products.
Here $A_{ij}^{(k)}$ denote the expansion coefficients of the 
massive OME $\langle j|O^i|j\rangle$ of a local twist-2 operator $O$ located at 
the line $i = Q, q, g$ with the 
external state $j = q,g$, where $Q$ denotes a massive quark, $q$ a massless quark, 
$g$ the gluon. The expansion coefficients of 
the massless Wilson coefficients $C_i^{l,(k)}$ depend on $N_F +1$ flavors and 
$\tilde{f}(N_F) = f(N_F)/N_F$. For the other massive Wilson coefficients see
Refs.~\cite{Buza:1995ie, Bierenbaum:2009mv}.
The massive OMEs have been calculated to three-loop order in the 
unpolarized cases in Refs.~\cite{Ablinger:2010ty,Ablinger:2014vwa,
Ablinger:2014nga,Behring:2014eya,Kawamura:2012cr,
Ablinger:2023ahe,Ablinger:2024xtt}
and in the polarized cases  
in Refs.~\cite{Ablinger:2014vwa,Blumlein:2021xlc,Ablinger:2019etw,Ablinger:2023ahe,
Ablinger:2024xtt}.\footnote{The one-- and two--loop corrections were calculated in 
Refs.~\cite{Witten:1975bh,Babcock:1977fi,Shifman:1977yb,Leveille:1978px,
Gluck:1980cp,Watson:1981ce} and \cite{Laenen:1992zk,Laenen:1992xs,Buza:1995ie,Bierenbaum:2007qe,
Bierenbaum:2008yu,Buza:1996xr,Hekhorn:2018ywm,Bierenbaum:2022biv}.}

Beyond the massive OMEs contributing
to the massive Wilson coefficients, there are also others, $(\Delta) A_{gq,Q}$ and 
$(\Delta) A_{gg,Q}$, playing a role in the variable flavor number 
scheme~\cite{Buza:1996wv,Behring:2014eya,Ablinger:2025joi,Ablinger:2014lka,
Ablinger:2022wbb,Behring:2021asx}. 
     
It has been shown in Ref.~\cite{Buza:1995ie} that for the structure function
$F_2(x,Q^2)$ the heavy-flavor corrections in the asymptotic region $Q^2 \gg m_Q^2$ agree with the complete 
result at the level of $O(1\%)$ if 
$Q^2/m_Q^2 > 10$ at two-loop order. This cut also safely removes the higher-twist effects,
cf.~Refs.~\cite{Blumlein:2008kz,Blumlein:2012se,Alekhin:2012ig}.
The structure function $F_2(x,Q^2)$ consists of the massless part, $F_2^{\rm light}(x,Q^2)$ for the 
light flavors, and the massive part, $F_2^{\rm heavy}(x,Q^2)$, 
where the Wilson coefficients have either 
a charm or a bottom quark contribution.\footnote{Here we calculate the 
`extrinsic' heavy-flavor contributions, but not intrinsic heavy-flavor 
corrections, as suggested by a Fock-space formalism in the 
infinite-momentum frame \cite{Brodsky:1980pb,Blumlein:2015qcn}.}
%---------------------------------------------------------------------------------------------------------------
\begin{eqnarray}
F_2(x,Q^2) = F_2^{\rm light}(x,Q^2)
+ F_2^{c}(x,Q^2)
+ F_2^{b}(x,Q^2).
\end{eqnarray}
%---------------------------------------------------------------------------------------------------------------
In our illustrations, we work in the fixed flavor number scheme for $N_F =  3$ light 
flavors with charges $e_k$. The N$^k$LO contribution to $F_2^\text{total}$ combines 
$F_2^\text{light}$ contributions through $O(a_s^k)$ and $F_2^\text{heavy}$ contributions 
through $O(a_s^{k+1})$ for $k = 0,1,2$.
The formal structures are the same for the structure functions
$F_2(x,Q^2)$ and $g_1(x,Q^2)$, and are obtained by substituting the unpolarized building 
blocks with the polarized ones. For $g_1(x,Q^2)$ the prefactor $x$ in 
Eqs.~(\ref{eq:SF1x},\ref{eq:SF2x}) has to be replaced by $1/2$.

The light-flavor part of $F_2$ is given by \cite{Vermaseren:2005qc,Blumlein:2022gpp}
%---------------------------------------------------------------------------------------------------------------
\begin{eqnarray}
F_2^{\rm light}(x,Q^2) &=& x \sum_{k=1}^{N_F} e_k^2 \Biggl\{\frac{1}{N_F}\Biggl[
C_{2,q}^{\rm S} \left(x, \frac{Q^2}{\mu^2}\right) \otimes \Sigma(x,\mu^2) 
%\nonumber \\ && 
+
C_{2,g}^{\rm S} \left(x, \frac{Q^2}{\mu^2}\right) \otimes G(x,\mu^2) \Biggr]
\nonumber\\ &&
+ C_{2,q}^{\rm NS}
\left(x, \frac{Q^2}{\mu^2}\right) \otimes  \Delta_k(x,\mu^2)\Biggr\},
\label{eq:SF1x}
\end{eqnarray} 
%---------------------------------------------------------------------------------------------------------------
where $\Delta_k(x,\mu^2), \Sigma(x,\mu^2)$ and $G(x,\mu^2)$ denote the non-singlet, 
singlet and gluon distribution with
%---------------------------------------------------------------------------------------------------------------
\begin{eqnarray}
\Delta_k(x,\mu^2) &=& f_k(x,\mu^2) + {f}_{\bar{k}}(x,\mu^2) - \frac{1}{N_F} \Sigma(x,\mu^2), \\
\Sigma(x,\mu^2)   &=& \sum_{k=1}^{N_F} [f_k(x,\mu^2) + {f}_{\bar{k}}(x,\mu^2)], 
\end{eqnarray} 
%---------------------------------------------------------------------------------------------------------------
with $f_{k,(\bar{k})}$ the quark and antiquark number densities,
and $\mu$ denotes the factorization scale, which we set equal to the renormalization scale.
The unpolarized massless Wilson coefficients $C_{2,i}^{j}$ 
were calculated to three-loop order in Refs.~\cite{Vermaseren:2005qc,Blumlein:2022gpp} 
and the polarized ones $\Delta C_{1,i}^{j}$ in 
Ref.~\cite{Blumlein:2022gpp}. The heavy-flavor part is given 
by, cf.~Ref.~\cite{Bierenbaum:2009mv},
%---------------------------------------------------------------------------------------------------------------
\begin{eqnarray}
\lefteqn{F_2^Q(x,Q^2) =} 
\nonumber\\ &&
x \Biggl\{\sum_{k=1}^{N_F} e_k^2\Biggl[
L_{2,q}^{\rm NS}(x) \otimes (f_k(x,\mu^2) + f_{\bar{k}}(x,\mu^2))
%\nonumber\\ &&
+ 
\frac{1}{N_F}\left({L}_{2,q}^{\rm PS}(x)
\otimes \Sigma(x,\mu^2)
+ {L}_{2,g}^{\rm S}(x)
\otimes G(x,\mu^2) \right)\Biggr]
\nonumber \\ &&
+ e_Q^2 \Biggl[
H_{2,q}^{\rm PS}(x)
\otimes \Sigma(x,\mu^2)
+ H_{2,g}^{\rm S}(x)
\otimes G(x,\mu^2)
\Biggr]\Biggr\}.
\label{eq:SF2x}
\end{eqnarray}
%------------------------------------------------------------------------------------------------
It contains both flavor non-singlet and singlet contributions.\footnote{For the study of 
pure non-singlet structure functions, see Ref.~\cite{Blumlein:2021lmf}. There the 
heavy-flavor corrections are
small.}
Here we have suppressed the dependence on $Q^2, m_Q^2$ and $\mu^2$
in the heavy-flavor Wilson coefficients $H_i$ and $L_i$, and $\otimes$ denotes the Mellin 
convolution,\footnote{For the evaluation of the convolution integrals in the numerical 
illustrations we use the integrator 
of Ref.~\cite{INT}.}
%------------------------------------------------------------------------------------------------
\begin{eqnarray}
A(x) \otimes B(x) = \int_0^1 dx_1 \int_0^1 dx_2 \delta(x - x_1 x_2) A(x_1) B(x_2).
\label{eq:MEL}
\end{eqnarray}
%------------------------------------------------------------------------------------------------
The following convolutions contribute,  cf.~Eqs.~(256--258) of 
Ref.~\cite{Bierenbaum:2022biv},
%----------------------------------------------------------------------------------------------------------------
  \begin{eqnarray}
\label{eq:con1}
  \left(\left[\frac{f}{1-x}\right]_+ \otimes g\right)(x) &=& \int_x^1
  dz \frac{f(z)}{1-z} \left[\frac{1}{z} g\left(\frac{x}{z}\right) - g(x) \right]
-
g(x) \int_0^x dz
  \frac{f(z)}{1-z}, \\
%---
\label{eq:con2}
(h \otimes g)(x) &=& \int_x^1 \frac{dz}{z} h(z) g\left(\frac{x}{z}\right),
\\
\label{eq:con3}
(\delta(1-x) \otimes g)(x) &=&
  g(x).
\end{eqnarray}
%----------------------------------------------------------------------------------------------------------------
Here $f(x)$ and $g(x)$ are functions with support $x \in ]0,1[$.
%------------------------------------------------------------------------------------------$
\begin{figure}[H]\centering
\includegraphics[width=0.8\textwidth]{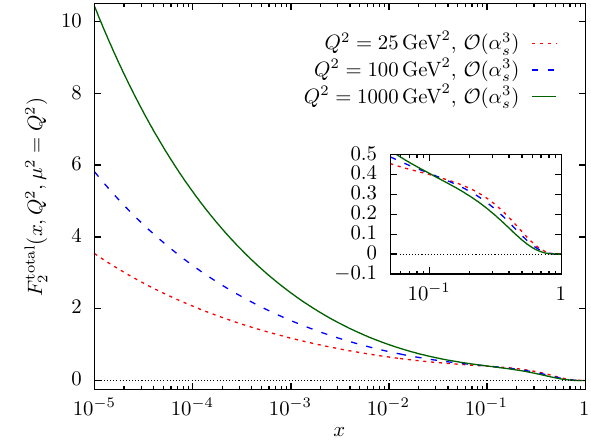}
\caption[]{\sf
The structure function $F_2(x,Q^2)$ at NNLO.}
\label{fig:1}
\end{figure}
%------------------------------------------------------------------------------------------$

In Mellin space, the massive Wilson coefficients are polynomials of the expansion coefficients of
the massive OMEs $(\Delta) A_{ij}$ and expansion coefficients of the massless Wilson coefficients, see 
Refs.~\cite{Bierenbaum:2009mv,Behring:2014eya}.
The massless Wilson coefficients to three-loop order can all be expressed in terms of either harmonic 
sums \cite{Vermaseren:1998uu,Blumlein:1998if} in Mellin space or harmonic polylogarithms 
\cite{Remiddi:1999ew} in $x$-space. 
For the massive Wilson coefficients, this only applies to the $N_F$-contributions 
\cite{Ablinger:2010ty} and to
$(\Delta) L_q^{\rm NS}, (\Delta) L_q^{\rm PS}$ and  $(\Delta) L_g^{\rm S}$. 
The pure-singlet Wilson 
coefficients $(\Delta) H_q^{\rm PS}$
depend on generalized harmonic sums \cite{Moch:2001zr,Ablinger:2013cf} in Mellin space and can be 
cast 
into harmonic 
polylogarithms with changed argument in $x$-space. The gluonic Wilson coefficients $(\Delta)H_g^{\rm S}$
receive also finite binomial sum \cite{Ablinger:2014bra} and higher transcendental function 
contributions
in Mellin space. They contain square-root valued iterated integrals \cite{Ablinger:2014bra}
 and $_2F_1$-solutions 
\cite{Ablinger:2017bjx} in 
$x$-space.\footnote{For surveys on the different mathematical structures and the used computer 
algebraic methods see Refs.~\cite{Blumlein:2018cms,Behring:2023rlq}.}  
Since we calculate the heavy-flavor corrections to inclusive structure functions, there 
are also contributions with massless final states, but virtual heavy-flavor corrections.

The paper is organized as follows. In Section~\ref{sec:2} we compute the 
single-mass heavy-quark corrections for 
the unpolarized structure function $F_2(x,Q^2)$ and in Section~\ref{sec:3} those for the 
polarized structure function $g_1(x,Q^2)$ to three-loop order. The details of the 
fast and precise numerical
code provided, which is well suited for QCD-analyses of hard-scattering data, are 
described in Section~\ref{sec:4} and the conclusions are contained in 
Section~\ref{sec:5}.
%- }}}
%- {{{ F2:

\section{The Structure Function \boldmath $F_2(x,Q^2)$} 
\label{sec:2}

\vspace*{1mm}
\noindent
In the unpolarized case, we 
use the parton distribution functions (PDFs) of Ref.~\cite{Alekhin:2017kpj}
for $N_F = 3$.\footnote{From LHAPDF \cite{LHAPDF}.} 
%------------------------------------------------------------------------
\begin{figure}[H]\centering
\includegraphics[width=0.80\textwidth]{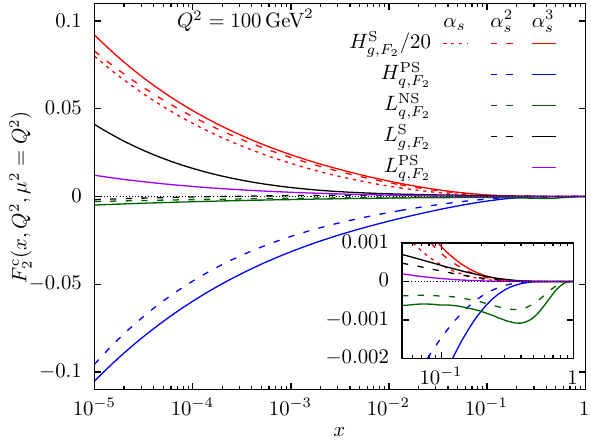}
\caption{\sf The different heavy-flavor contributions due to charm  quarks to the structure
function
$F_2(x,Q^2)$ by the Wilson coefficients  $H_g^{\rm S}, H_q^{\rm PS}, L_q^{\rm NS}, L_g^{\rm 
S}$ and
$L_g^{\rm PS}$ at $Q^2 = 100~\GeV^2$ at different orders in the strong coupling constant 
up to $O(a_s), O(a_s^2)$ and $O(a_s^3)$.} 
\label{fig:2}
\end{figure}  
%------------------------------------------------------------------------

For the values of $\alpha_s$ in the 
Wilson coefficients, we use flavor matching
with the values $\alpha_s(25~\GeV^2)   = 0.2020,
\alpha_s(10^2~\GeV^2) = 0.1706$, and $\alpha_s(10^3~\GeV^2) = 0.1359$, corresponding to 
$\alpha_s(M_Z^2) = 0.1147$, consistent with the PDF fits.
The massive Wilson coefficients were calculated referring to on-shell
heavy-quark masses. We use the following values
%------------------------------------------------------------------------------------------------
\begin{eqnarray}
m_c = 1.59~\GeV,~~~~~{\it m_b} = 4.78~\GeV,
\end{eqnarray}
%------------------------------------------------------------------------------------------------
cf.~Refs.~\cite{Alekhin:2012vu,Agashe:2014kda}.

In Figure~\ref{fig:1}, we illustrate the prediction for the 
total structure function $F_2^{\rm total}(x,Q^2)$, including the single-mass charm and 
bottom quark contributions. 
In the region $x \sim 10^{-5}$,  it rises from $Q^2 = 25~\GeV^2$ to $Q^2 = 1000~\GeV^2$ 
from values of $\sim 3.5$ to 10.5.

Figure~\ref{fig:2} shows the contributions of the different heavy-flavor Wilson 
coefficients to $F_2^c$ at $Q^2 = 100~\GeV^2$. The different lines illustrate also
the contributions from $O(a_s)$ to $O(a_s^3)$ to the structure function. 
The largest contribution is due to the gluonic Wilson 
coefficient
$H_g^{\rm S}$ which was scaled down by a factor of 20 for better readability. 
The 
corrections are positive and grow with the order in $a_s$. It is followed in size by a 
negative contribution from the pure-singlet Wilson coefficient $H_q^{\rm 
PS}$, which starts at $O(a_s^2)$. 
In the large-$x$ region, the non-singlet Wilson coefficient $L_q^{\rm NS}$ yields the 
dominant part.

%------------------------------------------------------------------------
\begin{figure}[H]\centering
\includegraphics[width=0.8\textwidth]{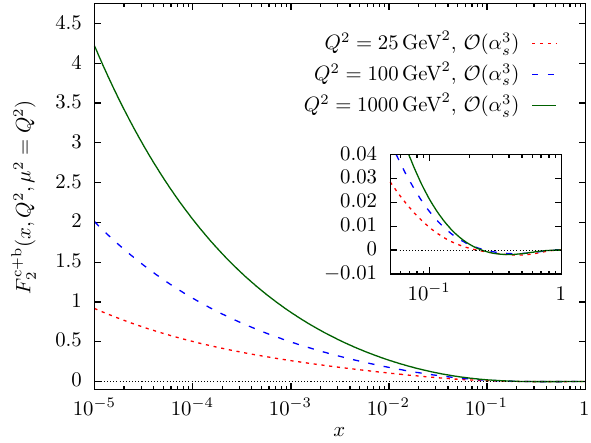}  
\caption{\sf The heavy-flavor contributions due to charm and bottom quarks to the
structure function $F_2(x,Q^2)$ at  NNLO.}
\label{fig:3}
\end{figure}  
%------------------------------------------------------------------------

Around $x = 0.1$ there is a turning point above which the structure 
function takes lower values for larger values of $Q^2$. 
However, the corrections compared to $F_2^{c,b}$ stay rather small. 
The corrections can become negative since virtual effects of heavy quarks with massless final 
states are contained in $F_{2}^{c}$ and $F_{2}^{b}$ and some Wilson coefficients start at
$O(a_s^2)$ only.
Also the contributions of $L_g^{\rm S}$ and $L_g^{\rm PS}$ are of importance 
at the level of accuracies of $O(1\%)$. $L_g^{\rm S}$ is tiny at $O(a_s^2)$, but receives a 
relative large correction at $O(a_s^3)$. A part of Wilson coefficients emerging at higher 
order in the coupling constant $a_s$ can be negative as long as the unpolarized structure 
function remains positive.

In Figure~\ref{fig:3} the heavy-flavor corrections to $F_2(x,Q^2)$ are shown at NNLO. At $x = 
10^{-5}$, they grow from 1 to $\sim 4$ for $Q^2 = 25~\GeV^2$ to $Q^2 = 1000~\GeV^2$. In 
the
large-$x$ region the corrections are negative.

\begin{figure}[H]\centering
\includegraphics[width=0.80\textwidth]{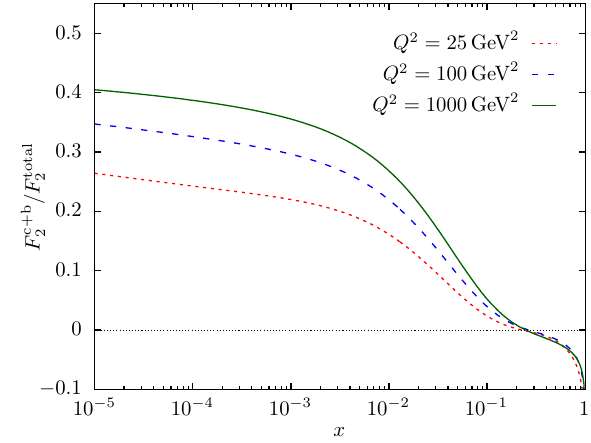}  
\caption{\sf The ratio of the heavy-flavor contributions due to charm and bottom quarks 
to the structure function $F_2(x,Q^2)$ at NNLO.}
\label{fig:4}
\end{figure}

The relative single-mass charm and bottom contributions 
to $F_2(x,Q^2)$ are shown in Figure~\ref{fig:4},
as the ratio
%------------------------------------------------------------------------------------------------
\begin{eqnarray}
\label{eq:rat}
R = \frac{F_2^{c} +F_2^{b}}{F_2^{\rm light} +F _2^{c} +F_2^{b}},
\end{eqnarray}
%------------------------------------------------------------------------------------------------
at NNLO. In the small-$x$ region
of $x \sim 10^{-5}$, the fractions of the heavy-flavor corrections evolve from 
$Q^2 = 25~\GeV^2$ to $Q^2 = 1000~\GeV^2$ from 26~\% to 41~\%.
At $x \sim 0.2$, the heavy-flavor corrections become zero, which is due to a combination of real and 
virtual corrections at NLO and NNLO. In the very-large-$x$ region,
the heavy-flavor contributions at NNLO become negative and as large as  $-10\%$,
but the structure function $F_2$ itself remains non-negative.
It is clearly seen that the scaling violations of the massive contributions are rather different
compared to those in the massless case up to high values of $Q^2$.
%------------------------------------------------------------------------
%------------------------------------------------------------------------
%- }}}
%- {{{ g1:

\section{The Structure Function \boldmath $g_1(x,Q^2)$} 
\label{sec:3}

\vspace*{1mm}
\noindent
In the polarized case the 
calculation is performed in the Larin scheme \cite{Larin:1993tq}. Both for 
the massive 
three-loop OMEs and the massless three-loop Wilson coefficients, the scheme transformations to the 
$\overline{\rm MS}$ scheme are not yet known. However, if all contributing quantities, 
including also the
%------------------------------------------------------------------------------------------$
\begin{figure}[t]\centering
\includegraphics[width=0.80\textwidth]{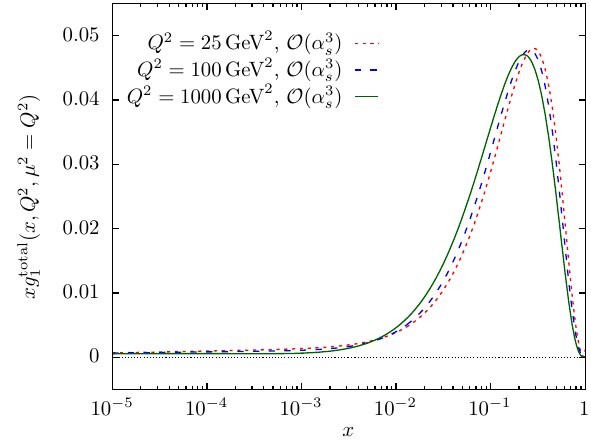}
\caption[]{\sf
The structure function $g_1(x,Q^2)$ at NNLO.}
\label{fig:5}
\end{figure}
%------------------------------------------------------------------------------------------$

\vspace*{-10mm}
%------------------------------------------------------------------------
\begin{figure}[H]\centering
\includegraphics[width=0.80\textwidth]{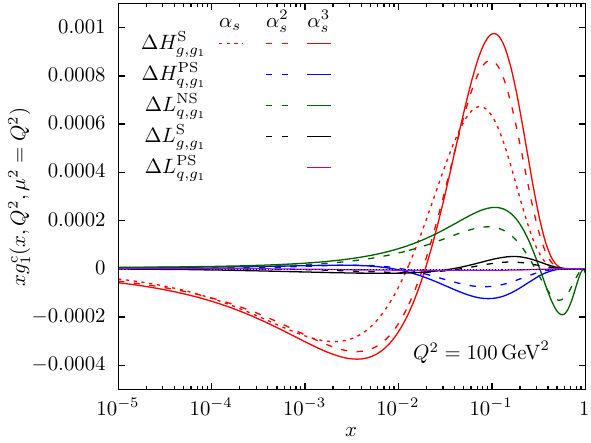}
\caption{\sf The different heavy-flavor contributions due to charm  quarks to the structure
function
$g_1(x,Q^2)$ due to the Wilson coefficients  $\Delta H_g^{\rm S}, \Delta H_q^{\rm PS}, \Delta L_q^{\rm 
NS}, \Delta L_g^{\rm S}$ and
$\Delta L_g^{\rm PS}$ at $Q^2 = 100~\GeV^2$ in the strong coupling constant   
up to $O(a_s), O(a_s^2)$ and $O(a_s^3)$.}
\label{fig:6}
\end{figure}  
%------------------------------------------------------------------------

\noindent
parton distribution functions, are used in this scheme, $g_1(x,Q^2)$, as a scheme independent 
quantity, can still be assembled. 

For the process of deep-inelastic scattering, the Larin scheme is a 
consistent scheme, given the number of contributing $\gamma_5$ Dirac matrices.
A set of Larin-scheme parton distribution functions has been generated in 
Ref.~\cite{Blumlein:2024euz} and is used in the following illustrations. 
The massless and 
massive two-loop corrections for the structure function $g_1(x,Q^2)$ were 
calculated in Ref.~\cite{Bierenbaum:2022biv}.
%------------------------------------------------------------------------------------------------

In Figure~\ref{fig:5}, we illustrate the scale evolution of the complete 
structure function $g_1^{\rm total}(x,Q^2)$ from $Q^2 = 25~\GeV^2$ to $Q^2 = 
1000~\GeV^2$ at three-loop order. 
With growing values of $Q^2$, $xg_1^{\rm total}(x,Q^2)$ grows below $x \sim 0.2$ and 
depletes at larger values of $x$. Also below $x \sim 10^{-2}$ 
the order in the scales reverts.
The structure functions $g_1^{p,d}(x,Q^2)$ are positive 
within errors, cf.~Ref.~\cite{ParticleDataGroup:2024cfk}.

Figure~\ref{fig:6} illustrates the contributions of the different heavy-flavor Wilson 
coefficients to $g_1^c$ for $Q^2 = 100~\GeV^2$, also with their contributions 
up to a given order in $a_s$. One clearly sees the importance of the 
higher-order corrections. The largest contribution is due to the 
gluonic Wilson coefficient $\Delta H_g^{\rm S}$. As has been discussed in 
Refs.~\cite{Buza:1996xr,Bierenbaum:2022biv}, the first moment of $\Delta H_g^{\rm 
S}(x,Q^2)$ vanishes, 
%------------------------------------------------------------------------------------------------
\begin{eqnarray}
\label{eq:SR}
\int_0^1 dx \Delta H_{g}^{\rm S}(x,Q^2) = 0.
\end{eqnarray}
%------------------------------------------------------------------------------------------------
%------------------------------------------------------------------------
\begin{figure}[H]\centering
\includegraphics[width=0.80\textwidth]{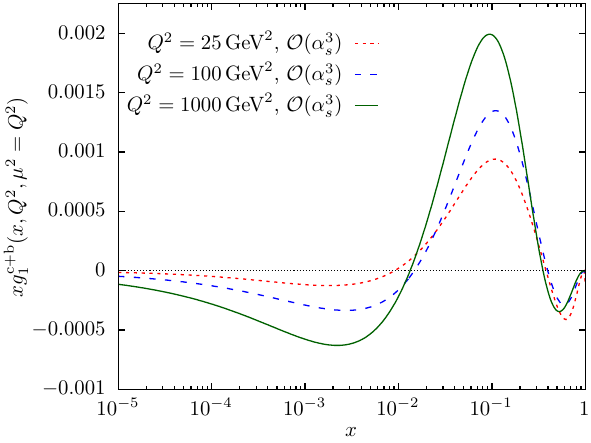}  
\caption{\sf The heavy-flavor contributions due to charm and bottom quarks to the structure function 
$g_1(x,Q^2)$ at NNLO.}
\label{fig:7}
\end{figure}  
%------------------------------------------------------------------------

This is now also found at three-loop order. Mainly due to this, the oscillating structure 
in 
Figure~\ref{fig:6} is implied.
Next in size are the flavor non-singlet and pure-singlet Wilson coefficients, 
$\Delta L_q^{\rm NS}$ and $\Delta H_q^{\rm PS}$. In the small-$x$ region, the former  is 
positive and the latter negative. $\Delta L_q^{\rm NS}$ turns negative above $x \sim 
0.2$ and dominates the large-$x$ region. The contributions due to 
$\Delta L_g^{\rm S}$ and $\Delta L_g^{\rm PS}$ are smaller but not negligible.

Figure~\ref{fig:7} shows the NNLO heavy-flavor contributions to $xg_1(x,Q^2)$. They grow
in size with $Q^2$. Below $x \sim 0.01$ they are negative and turn positive until at $x \sim
0.3$ they turn negative again.

\begin{figure}[H]\centering
\includegraphics[width=0.80\textwidth]{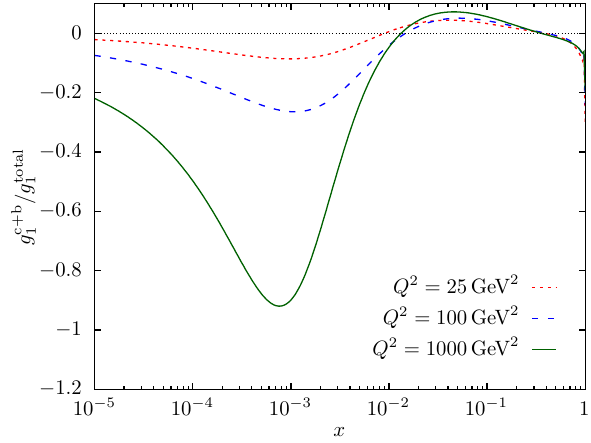}  
\caption{\sf The ratio of the heavy-flavor contributions due to charm and bottom quarks 
to the structure function $g_1(x,Q^2)$ at NNLO.}
\label{fig:8}
\end{figure}

In Figure~\ref{fig:8}, we show the ratio Eq.~(\ref{eq:rat}) of the charm and bottom contributions 
to the 
whole structure function $g_1(x,Q^2)$ at NNLO for $Q^2 = 25, 100$ and $1000~\GeV^2$. 
In the small-$x$ region, the corrections are negative. The largest 
negative values are about $-10~\%, -25~\%$ and $-95~\%$ from $Q^2 = 25, 100$ to 
$1000~\GeV^2$. These large corrections occur in a region 
in which $xg_1(x,Q^2)$ itself is rather small, see Figure~\ref{fig:5}.
Around $x = 10^{-2}$, the ratio turns positive. At $x = 0.3$ it turns 
negative again, as in the unpolarized case. The complete structure function 
$g_1(x,Q^2)$
remains positive. The strong negative correction in the small-$x$ region is mainly 
induced by 
the sum rule of the vanishing first moment of $\Delta H_g^{\rm S}$. 

%------------------------------------------------------------------------
%------------------------------------------------------------------------

%- }}}

\section{The numerical code}
\label{sec:4}
  
\vspace*{1mm}
\noindent
We provide the {\tt Fortran}-library {\tt WILS3HQ} of the massive Wilson 
coefficients to three-loop order as a fast and precise numerical implementation. It is 
well suited for QCD analyses of deep-inelastic structure functions in the
unpolarized and polarized case. The Wilson coefficients are represented as polynomials 
in the parameters $a_s, N_F$, $\ln(Q^2/\mu^2), \ln(m^2/\mu^2)$ and {\tt flav}, see below.
Also the $x$-dependence is described by elementary functions and polynomial interpolations.  
The order 
in the coupling constant is selected by the parameter {\tt IO} = 1,2,3. 
Both for the structure functions $F_2$ and $g_1$ five different Wilson 
contribute. In the non-singlet cases $L_{q,2(1)}^{\rm NS}$ there are two 
distribution valued contributions 
in addition to the regular part. Likewise, for the Wilson coefficients $H_{2(1),q}^{\rm 
PS}$ and $L_{2(1),q}^{\rm PS}$, $\delta(1-x)$-terms contribute.
The regular terms are represented by suitably deep analytic  small- and large-$x$ 
expansions and grid interpolations for the remainder parts for $x \in [0,1]$, optimized 
for  $x \geq x_{\rm  min} = 10^{-6}$. The grid interpolations, performed by third order 
splines \cite{SPLINE}, are based on 501 points.\footnote{We thank S. Kumano 
and M. Miyama of the AAC-collaboration for allowing us to use their 
interpolation routines.} The provided spline representation can be addressed directly by 
setting 
{\tt IDAT = 1}.
The $+$-and $\delta$-distribution parts are given in analytic 
form.
The representations need only elementary functions and we used code optimization 
\cite{Ruijl:2017dtg}.
The Mellin convolutions with the parton distributions have to be performed in the 
respective implementations by the users.
The data-set initialization is done by the routines {\tt INIDAT1 ... 10}, where the order
of the Wilson coefficients is {\tt H2gS, H2qPS, L2qNS, L2gS, L2qPS; HG1gS, HG1qPS, LG1qNS, 
LG1gS, LG1qPS}. 

For the grid-files we use the following name convention, illustrated by the example
%------------------------------------------------------------------------------------------------
\begin{eqnarray}
\label{eq:nam}
{\tt L2qPS30011.dat}.
\end{eqnarray}
%------------------------------------------------------------------------------------------------
The name starts with the name of the Wilson coefficient. The number sequence denotes the 
order in $a_s$, the power of $\ln(m^2/\mu^2)$, the power of $\ln(Q^2/\mu^2)$, the 
presence of the factor $N_F$, the presence of the factor $\tt flav$. The set in 
Eq.~(\ref{eq:nam}) corresponds to the grid for the massive Wilson coefficient 
$L_{2q}^{\rm PS (3)}$ of the structure function $F_2$, proportional to the factor $a_s^3 
N_F {\tt flav}$.
\begin{table}[t]
  \renewcommand{\arraystretch}{1.5}
  \centering
  \begin{tabular}{|c|c|c|}
      \hline
      & charm & bottom \\ 
      \hline
      $L_q^{{\rm NS}}$ & $2$ & $-1$ \\ 
      \hline
      $L_q^{{\rm PS}}$ & $-\frac{2}{5}$ & $\frac{2}{7}$ \\
      \hline
      $L_g^{{\rm S}}$  & $ \frac{2}{5}$ & $\frac{1}{7}$ \\ 
      \hline
      $H_q^{{\rm PS}}$ & $-\frac{2}{5}$ & $\frac{2}{7}$ \\ 
      \hline
      $H_g^{{\rm S}}$  & $ \frac{2}{5}$ & $\frac{1}{7}$  \\
      \hline
  \end{tabular}
  \caption{\sf Numerical values of the factor {\tt flav} for the perturbative 
  generation of a heavy charm or bottom quark.}
  \label{tab::flav}
  \renewcommand{\arraystretch}{1}
\end{table}

%------------------------------------------------------------------------------------------------
For the charm quark we find the following formulae for the flavor factors\footnote{These qunatities are related to the numbers ${\sf w}_2$ and 
${\sf w}_3$ in Refs.~\cite{Vermaseren:2005qc,Blumlein:2022gpp}.} 

\begin{alignat}{3}
  L_q^{{\rm NS}}: &~~{\tt flav} &~~=~~& 3 e_c,\\
  L_q^{{\rm PS}}: &~~{\tt flav} &~~=~~& - \frac{3 e_c (2-e_c+3 e_c^2)}{4 (2 + 3 
e_c^2)},\\
  L_g^{{\rm S}} : &~~{\tt flav} &~~=~~& \frac{3 e_c^2}{2 + 3 e_c^2},\\
  H_q^{{\rm PS}}: &~~{\tt flav} &~~=~~& - \frac{3 e_c (2-e_c+3 e_c^2)}{4 (2 + 3 
e_c^2)},\\
  H_g^{{\rm S}} : &~~{\tt flav} &~~=~~& \frac{3 e_c^2}{2 + 3 e_c^2},
\end{alignat}
with $e_c$ the electric charge of the charm quark and we already inserted 
the charges of the three light quark flavors. The corresponding numbers in the case of 
bottom are obtained by replacing $e_c$ with $e_b$.
The flavor factors for $L_q^{\rm PS}$ and $H_q^{\rm PS}$ as well as 
$L_g^{\rm S}$ and $H_g^{\rm S}$ are the same, since all flavor factors 
vanish when considering only the three light quark charges.
These flavor factors result from the pieces of the three-loop massless Wilson 
coefficients in the asymptotic massive Wilson coefficients.

The regular parts of the unpolarized Wilson coefficients are denoted by {\tt REG...f}, 
as,~e.g., {\tt REGL2qNS.f} in the flavor non-singlet Wilson coefficient for the structure 
function $F_2$ and, correspondingly, with {\tt REGLG1qNS.f} for the structure function $g_1$. 
The routines of the $+$-distribution 
parts are named by {\tt PLU...f}, and the coefficients of the $\delta(1-x)$ terms by {\tt DEL...f}. 
The argument list is
%------------------------------------------------------------------------------------------------
\begin{eqnarray}
{\tt (IO,X,as,LM,LQ,NF,flav)},
\end{eqnarray}
%------------------------------------------------------------------------------------------------
where ${\tt as} = a_s(\mu^2), {\tt LM} = \ln(m^2/\mu^2), {\tt LQ} = \ln(Q^2/\mu^2), {\tt NF} 
= N_F, {\tt X} = x$, the Bjorken variable. In the case of the $\delta(1-x)$ terms {\tt X} is not part 
of 
the argument list. For the regular parts there are the auxiliary 
functions {\tt SX....f}, {\tt LX....f}, and {\tt GR....f}, for the small- and large-$x$
expansions, and the grid part. 

The different functions  of the library can 
be compiled with {\tt gfortran}. We provide ten programs {\tt MAIN1.f} to {\tt MAIN10.f}
to test the set-up of the respective Wilson coefficients by comparing with numerical results
obtained by our initial (semi)-analytic calculation, for some specific choice of parameters.

We add the numerical integration routine {\tt DAIND} of Ref.~\cite{DAIND} to perform the 
convolution integrals with the parton densities for the convenience of the user.
The license conditions to use the library {\tt WILS3HQ} are to quote the papers
Refs.~\cite{Bierenbaum:2009mv,
Ablinger:2010ty,
Ablinger:2014vwa,
Ablinger:2014nga,
Behring:2014eya,
Ablinger:2019etw,
Blumlein:2021xlc,
Ablinger:2023ahe,
Ablinger:2024xtt} and the present paper, in which the single-mass
three-loop corrections were calculated.

%- {{{ Concl.:

\section{Conclusions}
\label{sec:5}
  
\vspace*{1mm}
\noindent
In this paper, we present for the first time numerical results for the single-mass 
three-loop heavy-flavor
corrections to the unpolarized structure function $F_2(x,Q^2)$ and the polarized structure 
function $g_1(x,Q^2)$ at twist-2. The asymptotic analytic results on the 
heavy-flavor Wilson coefficients are correlated to the kinematic region in which the 
higher-twist contributions
can be neglected. Our numerical results illustrate the impact of the different Wilson 
coefficients 
and show the changes in the structure functions going from the 
$O(a_s)$ to the $O(a_s^3)$ corrections.
In the unpolarized case, the corrections are large in the small-$x$ region, which is also 
observed in the polarized case. Here, however, the large corrections emerge in a region
where the structure function is small, unlike in the former case.
The present results are an important step to allow for consistent QCD analyses of 
the deep-inelastic World data at the level of NNLO, given the 
partial results used before for combined singlet/non-singlet data analyses.
Our new results  will allow to improve the accuracy of the strong
coupling constant $a_s(M_Z^2)$, the PDFs and the charm mass, extracted from the 
deep-inelastic World data. 

A fast and numerically precise public {\tt Fortran} code for the calculation of the 
asymptotic single-mass heavy-flavor Wilson coefficients, {\tt WILS3HQ}, 
for the use in experimental data analyses, is provided
as an ancillary file to this paper.
%- }}}

\vspace*{5mm}
\noindent
{\bf Acknowledgments.}  
We would like to thank I.~Bierenbaum, A.~Goedicke (n\'e Hasselhuhn), S.~Klein, C.~Raab, M.~Round, 
M.~Saragnese and F.~Wi\ss{}brock for collaboration in earlier projects and thank P.~Marquard for 
discussions. This work was supported by the European Research Council (ERC)
under the European Union's Horizon 2020 research and innovation programme
grant agreement 101019620 (ERC Advanced Grant TOPUP), the UZH Postdoc Grant,
grant no.~[FK-24-115] and by the Austrian Science Fund (FWF) Grant-DOI 10.55776/P20347.
%-----------------------------------------------------------------------------------------------------

%-----------------------------------------------------------------------------------
\end{document}